\documentclass[nofootinbib,twocolumn,pre,superscriptaddress,10pt,aps,floatfix]{revtex4-2}
\usepackage[english]{babel}
\usepackage{graphicx}
\usepackage{amsmath}
\usepackage{amssymb}
\usepackage[colorlinks,linkcolor=blue,citecolor=blue,urlcolor=blue]{hyperref}
\usepackage{color}
\usepackage{soul}
\usepackage[normalem]{ulem}
\usepackage{xcolor}
\usepackage{makecell}
\usepackage{multirow}
\definecolor{dblue} {RGB}{28,130,185}
\let\oldaddcontentsline\addcontentsline
\newcommand{\stoptocentries}{\renewcommand{\addcontentsline}[3]{}}
\newcommand{\starttocentries}{\let\addcontentsline\oldaddcontentsline}
\definecolor{nred}{RGB}{224,0,0}
\definecolor{nblue}{RGB}{28,130,185}
\definecolor{darkgreen}{rgb}{0,0.60,.2}
\definecolor{pgreen}{rgb}{0,1.0,0.}

\newcommand\redsout{\bgroup\markoverwith{\textcolor{brown}{\rule[0.6ex]{4pt}{1.0pt}}}\ULon}
\newcommand{\mtc}[1]{\mathcal{#1}}
\setcellgapes{2pt}
\makegapedcells

\begin{document}
\title{Quadrupole gauge theory: anti-Higgs mechanism and elastic dual}

\author{Aleksander G\l{}\'odkowski}

\affiliation{Institute of Theoretical Physics, Faculty of Fundamental Problems of Technology, Wroc\l{}aw University of Science and Technology, 50-370 Wroc\l{}aw, Poland}
\author{Pawe\l{} Matus}

\affiliation{Max Planck Institute for the Physics of Complex Systems, N\"othnitzer Stra{\ss}e 38, 01187 Dresden, Germany}
\affiliation{W\"urzburg-Dresden Cluster of Excellence ct.qmat, 01187 Dresden, Germany}
\author{Francisco Peña-Benítez}

\affiliation{Institute of Theoretical Physics, Faculty of Fundamental Problems of Technology, Wroc\l{}aw University of Science and Technology, 50-370 Wroc\l{}aw, Poland}
\author{Lazaros Tsaloukidis}

\affiliation{Max Planck Institute for the Physics of Complex Systems, N\"othnitzer Stra{\ss}e 38, 01187 Dresden, Germany}
\affiliation{W\"urzburg-Dresden Cluster of Excellence ct.qmat, 01187 Dresden, Germany}

\begin{abstract}
Motivated by the duality between elasticity and fracton gauge theory, we study an extension of the gauge group that includes the quadrupole moment. Remarkably, we find that spontaneous breaking of the quadrupole symmetry increases the number of massless excitations. This result appears to challenge the well-established paradigm, according to which gauge fields acquire mass through the Higgs mechanism, and the would-be Goldstone bosons are rendered massive. We refer to this phenomenon as the anti-Higgs mechanism. Furthermore, we demonstrate that the quadrupole gauge theory is dual to an exotic class of elastic systems, which we dub \emph{incompressible crystals}. The anti-Higgs mechanism in the dual theory corresponds to a phase transition from the incompressible crystal state to an ordinary (compressible) solid. 
\end{abstract}
\maketitle
\stoptocentries
%================================================================================
{\it Introduction.}---Fundamental particles in the Standard model (with the exception of neutrinos) acquire their masses via the Higgs mechanism \cite{Anderson1963,Englert1964,Higgs1964,Guralnik1964}, which generalizes the phenomenon of spontaneous symmetry breaking (SSB) to gauge groups. The core idea of the Higgs mechanism is that gauge bosons gain mass through minimal coupling to a charged field that develops a vacuum expectation value (in a fixed gauge). In the case of a global symmetry, this would result in the appearance of a gapless Nambu-Goldstone mode; for a gauge symmetry, however, the analogue field usually called Stueckelberg gets ``eaten" by the gauge field. The Higgs mechanism provides a general theoretical framework for generating a gap in gauge theories, playing a foundational role in describing various physical phenomena, such as the Meissner effect in superconductors \cite{Anderson1963}, synthetic gauge fields in cold atom experiments \cite{PhysRevLett.111.115303}, and certain models of cosmological inflation \cite{Bezrukov:2007ep}, to name a few. 

While the SSB mechanism is well understood for systems with internal symmetries, the picture is less clear when they are not uniform and the generators do not commute with spatial translations; common examples of nonuniform symmetries include spacetime symmetries such as rotations and boosts. Currently, a complete understanding of SSB for nonuniform symmetries remains an open problem under active investigation \cite{Ivanov:1975zq,Low:2001bw,PhysRevD.89.085004,Klein:2017npd,Alberte:2020eil,Hirono_2024}. In this letter, we study a gauge theory with a nonuniform symmetry group that challenges the Higgs paradigm: in our system, the mechanism turns out to \textit{add} a massless mode instead of gapping out the gauge field.

The model under consideration is motivated by the physics of \textit{fractons}--a topic that has attracted attention from diverse areas in theoretical physics due to its unconventional properties, such as restricted mobility \cite{Vijay:2015mka,Vijay:2016phm,Pretko2017a,Pretko2017b}, UV/IR mixing \cite{Seiberg:2020bhn,Gorantla:2021bda}, and slowdown of hydrodynamic transport \cite{Gromov:2020yoc,Grosvenor:2021rrt,Glorioso:2021bif,Glodkowski:2022xje,Glorioso:2023chm,Glodkowski:2024ova}, among many others (see \cite{Nandkishore:2018sel,Pretko:2020cko,Grosvenor:2021hkn,Gromov:2022cxa} for a review). Fractons are most prominently realized as elementary charges in systems with \textit{multipole symmetries} associated with the conservation not only of the total charge but also some of its higher multipole moments \cite{Pretko2017a,Gromov2019}. Moreover, gauging the multipole symmetry group generally results in the emergence of exotic higher-rank gauge theories characterized by tensor gauge fields \cite{Pretko:2018jbi, Pena2023,Jain:2021ibh,Bidussi:2021nmp,hoyos2022}. 

Remarkably, an exact correspondence exists between certain classes of tensor gauge theories and the low-energy description of solids, known as the fracton-elasticity duality, where the elasticity theory is mapped onto the higher-rank gauge theory, with elastic defects serving as fractonic charges \cite{Pretko2018a,Pretko2018b,PhysRevLett.121.235301,PhysRevB.100.045119,Pretko2019,Surowka2020,Nguyen_2020,Vijay,hoyos2022,Nguyen_2024b,Surowka2024a,Surowka2024b}. A prominent example,  first observed by Pretko and Radzihovsky \cite{Pretko2018a}, is the duality between ordinary elastic solids in two dimensions and a symmetric tensor gauge theory that conserves both an abelian charge and its corresponding
dipole moment. This \textit{dipole gauge theory} can be systematically constructed by gauging the dipole symmetry group \cite{Pretko:2018jbi,Pena2023}.

In this letter, we generalize previous approaches and study the gauge theory of a quadrupole-conserving symmetry group. As we will argue below, such \textit{quadrupole gauge theory} is dual to an \textit{incompressible crystal}, which is a different phase of matter than ordinary solids hosting a single magnon-like transverse sound mode. Interestingly, ordinary elasticity theory is obtained after driving such an incompressible crystal through a phase transition, with the
symmetry-broken phase given precisely by the theory of Pretko and Radzihovsky \cite{Pretko2018a}.   

{\it Quadrupole gauge theory.}---We consider a system characterized by the nonuniform symmetry group with generators $P_a, Q^{(0)},Q^{(1)}_a$ and $Q^{(2)}$, which we refer to as momentum, monopole, dipole, and quadrupole, respectively. The algebra of the conserved charges---\textit{the quadrupole algebra}---contains the following non-trivial brackets: 
\begin{equation}\label{eq:algebra}
    [P_a, Q^{(1)}_b] = \delta_{ab} Q^{(0)}\,, \quad [P_a,   Q^{(2)}] = Q^{(1)}_a\,.
\end{equation}
The conservation of currents associated with the charges $Q^{(0)},Q^{(1)}_a$, and $Q^{(2)}$ is expressed as
\begin{equation}
    d \star j^{(0)} = 0\,, \quad  d \star j^{(1)}_a = 0\,, \quad d \star j^{(2)} = 0 \,,
\end{equation}
where $\star j$ is the Hodge dual of $j$ with respect to the Minkowski metric $\text{diag}(-,+,\dots,+)$ and $d$ is the exterior derivative. We adopt the convention with $\epsilon^{01\dots d}= -1$ and $\epsilon_{01\dots d}= 1$. Importantly, the quadrupole algebra \eqref{eq:algebra} implies that the dipole and quadrupole currents are nonuniform\footnote{Uniform currents are defined by the property that the momentum operator acts on them as a derivative \cite{Hirono_2024}.} 
\begin{equation}
    j_a^{(1)} = J_a^{(1)} - x_a j^{(0)}  \,, \quad   j^{(2)} =   J^{(2)} -  x_a  j_a^{(1)}  - \frac{x^2}{2} j^{(0)} \,,
\end{equation}
where $J_a^{(1)}$ and $J^{(2)}$ are uniform but not conserved. In order to gauge the quadrupole algebra\footnote{In \ref{app:coset_quadrupole} we provide a complementary derivation using the coset construction method. See also \cite{hoyos2022}.} we introduce the gauge fields $A^{(0)},  A_a^{(1)}$ and $A^{(2)}$, which couple to the uniform currents $j^{(0)}, J_a^{(1)}$ and $J^{(2)}$, respectively. The gauge fields are endowed with the following transformation properties under gauge transformations  
\begin{equation}\begin{split}
    \delta A^{(0)} &= -d\lambda^{(0)} - \lambda^{(1)}_a dx_a \,, \\
    \delta A^{(1)}_a &= -d \lambda^{(1)}_a - \lambda^{(2)} dx_a\,, \\
     \delta A^{(2)} &= -d\lambda^{(2)}\,.
    \end{split}
\end{equation}
The gauge-invariant structures of the theory are then given by 
\begin{equation} \label{eq:strengths}
\begin{split}
    F^{(0)} &= d A^{(0)} + dx_a \wedge A_a^{(1)} \,, \\
     F_a^{(1)} &= d A^{(1)}_a + dx_a \wedge A^{(2)} \,, \\
      F^{(2)} &= d A^{(2)}\,.
    \end{split}
\end{equation}
It is straightforward to verify the following Bianchi identities:
\begin{equation}
    d  F^{(0)}=dx_a  \wedge F^{(1)}_a\,, \quad 
      d  F^{(1)}_a  = dx_a  \wedge F^{(2)}\,, \quad
      d  F^{(2)} = 0\,.
\label{eq:bianchis}
 \end{equation}
 Furthermore, in order to also capture the Higgs phase of the theory, we introduce a Stueckelberg field $\psi^{(2)}$ associated with the quadrupole condensate, which transforms nonlinearly under the action of the quadrupole symmetry $\psi^{(2)} \rightarrow \psi^{(2)} + \lambda^{(2)}$. In the presence of $\psi^{(2)}$, there is one more gauge-invariant quantity $U = A^{(2)} + d \psi^{(2)}$. The Maxwell action is then constructed in terms of the gauge invariant objects. In particular, we consider the following action\footnote{Note that Eq.~\eqref{eq:action} is not the most general action compatible with the symmetries. However, for the sake of brevity, we consider a simplified action. This simplification does not affect our conclusions.}
\begin{equation}
\begin{split}
       S_{\mathrm{EM}}  = -\frac{1}{2}\int & \Big(  l^{-2} F^{(0)} \wedge \star F^{(0)} +   F_a^{(1)} \wedge \star F_a^{(1)}   \\
       &+   L^2 F^{(2)}\wedge \star F^{(2)} + \lambda^2 U \wedge \star U \Big) \,.
\end{split} \label{eq:action}
\end{equation}
The coefficients $l$ and $L$ have dimensions of length while $\lambda$ is dimensionless. The Higgs phase is characterized by $\lambda \neq 0$. Equations of motion for the gauge fields are found by varying \eqref{eq:action} with respect to the gauge fields and the Stueckelberg field:
\begin{equation}\begin{split}
    d \star F^{(0)}&=0\,, \quad 
     l^2 d \star F^{(1)}_a  = \star F^{(0)}  \wedge dx_a\,, \\
     L^2 d \star F^{(2)} &=  \star F^{(1)}_a \wedge dx_a +\lambda^2 \star U\,, \quad d \star U =0\,.
 \label{eq:maxwells}
  \end{split}
 \end{equation}
 Together, the system of equations (\ref{eq:bianchis}) and (\ref{eq:maxwells}) describes the dynamics of the quadrupole gauge theory. 
 
\textit{Anti-Higgs mechanism.}---The Maxwell action~(\ref{eq:action}) contains massive degrees of freedom associated with the fields $A^{(1)}_{a0}, A^{(1)}_{[ab]}=\frac{1}{2}\epsilon_{ab} A^{(1)}_{ab}$, and $A^{(2)}_\mu$, which appear without derivatives in the expressions in Eq.~(\ref{eq:strengths}). In order to access the low-energy structure of the theory, we integrate out these massive fields (see \ref{app:modes} for computational details). Interestingly, we find that in the unbroken phase ($\lambda =0$), integrating out $A^{(2)}_\mu$ also removes the trace of the dipole gauge field $A^{(1)}_{aa}$ from the action. This is in line with the gauge-invariance of the action: since $A^{(2)}_\mu$ and $A^{(1)}_{aa}$ are the only fields that transform under quadrupole gauge transformations, integrating out one automatically removes the other, preserving gauge invariance at low energies. Finally, after fixing the gauge $A^{(0)}_\mu=0$, we are left with the traceless symmetric component of $A^{(1)}_{ab}$, $A^{(1)}_{\langle ab\rangle}= \frac{1}{2}(A^{(1)}_{ab}+A^{(1)}_{ba}-\frac{2}{d} \delta_{ab} A_{cc})$, as the only gapless degree of freedom, with the corresponding electric field $E^{(1)}_{\langle ab\rangle}=-\partial_0 A^{(1)}_{\langle ab\rangle}$ obeying the Gauss's law constraint
$\partial_a\partial_b E^{(1)}_{\langle ab\rangle}=0$. 
Thus, there exists $2-1=1$ gapless mode in two dimensions and $5-1=4$ gapless modes in three dimensions. Curiously, through an explicit derivation of the effective low-energy action (see \ref{app:modes}) we find that the dispersion relation of the massless mode in two dimensions is quadratic ($\omega \sim k^2$).

We now proceed to describe the Higgs phase ($\lambda\neq0$). Following the Higgs paradigm, one would generally expect the number of massless excitations in the Higgs phase to be reduced. However, in quadrupole gauge theory, introducing the Stueckelberg field $\psi^{(2)}$ does not render $A^{(2)}_\mu$ massive because $A^{(2)}_\mu$ is already massive in the unbroken phase. At the same time, integrating out $A^{(2)}_\mu$ does not enforce $A^{(1)}_{aa}$ to disappear from the action, as there still exists a gauge-invariant combination, $\partial_\mu \psi^{(2)} + \frac{1}{d}\partial_\mu A_{aa}^{(1)}$. After fixing the gauge $A^{(0)}_\mu=\psi^{(2)}=0$, the remaining gapless degrees of freedom are the symmetric components of $A^{(1)}_{ab}$, $A^{(1)}_{(ab)}=\frac12 \left(A^{(1)}_{ab}+A^{(1)}_{ba}\right)$. They are constrained by Gauss's law $\partial_a \partial_b E^{(1)}_{(ab)}=0$, where $E^{(1)}_{(ab)}=-\partial_0  A^{(1)}_{(ab)}$. Evidently, Higgsing the quadrupole gauge symmetry changes the character of the low-energy theory from the traceless symmetric tensor gauge theory to the traceful one, increasing the number of gapless modes from 1 to 2 in two dimensions and from 4 to 5 in three dimensions. In this phase, all modes are linear.

{\it Incompressible crystals.}---In this section, we introduce incompressible solids as a class of systems characterized by the symmetry algebra \eqref{eq:magneticAlegebra}, and construct their low-energy effective field theory. As shown in the next section, the theory of incompressible elasticity admits a dual description in terms of the quadrupole gauge theory.

Let us consider a homogeneous and isotropic solid in $(2+1)$--dimensions. In the long-wavelength limit, an effective description of the crystal is given in terms of $d=2$ scalar fields $\Phi_a \equiv \Phi_a(x,t)$, which represent the comoving (Lagrangian) coordinates of the solid element located at point $x$ at time $t$. The dynamics of the solid respect the \textit{internal symmetries} associated with translational and rotational invariance in $\Phi$-space \cite{Son2005,Dubovsky:2005xd},
\begin{equation}\label{eq:internal}
    \Phi_a \rightarrow D_{a b}(\theta)\Phi_b+\xi_a\,, 
\end{equation}
where $ D\in SO(2)$ represents rotation by an angle $\theta$ and $\xi \in \mathbb{R}^2$ denotes a coordinate shift. The corresponding generators are denoted by $L$ and $T_a$, repectively. 
In addition, we impose a global symmetry $Q$ together with the algebra
\begin{equation}\label{eq:magneticAlegebra}
    [T_a\,, T_b] =  \epsilon_{ab}  Q\,, \quad [L\,, T_a] = \epsilon_{ab} T_b\,.
\end{equation} 
We will call a system with internal symmetries that obey Eq.~\eqref{eq:magneticAlegebra} an \textit{incompressible crystal}. This naming will be justified \textit{a posteriori}, as we will show that imposing these commutation relations enforces the density of lattice sites to remain constant at low energies -- see Eq. \eqref{eq:inv_higgs2} and the discussion below. 

The algebra \eqref{eq:magneticAlegebra} is reminiscent of the algebra of magnetic translations familiar in the context of a charged particle in a magnetic field \cite{PhysRevB.54.5334}. However, in this work, we assume that our system is time-reversal and parity invariant. One realization of a system exhibiting  noncommutative translations while preserving  time-reversal symmetry is given by a Wigner crystal interacting with a dynamical gauge field \cite{Du2024}. 

In order to realize the global symmetry $Q$, we introduce a scalar field $\Psi \equiv \Psi(x,t)$ such that $Q$ acts on $\Psi$ with a shift symmetry $\Psi \rightarrow \Psi + \alpha$. Furthermore, the noncommutative structure of the algebra~\eqref{eq:magneticAlegebra} implies that the field $\Psi$ transforms under the action of $T_a$ as $\Psi \rightarrow \Psi + \frac{1}{2}  \epsilon_{ab} \xi_a \Phi_b$. In addition, we introduce generators of the \textit{spacetime symmetries}: Hamiltonian $H$, momentum $P_i$, and angular momentum $J$, which act on the spacetime coordinates in the usual way: $t  \rightarrow t + c_0$, $
x_a \rightarrow x_a + c_a$, and $
x_a \rightarrow R_{a b}(\alpha) x_b$, respectively.

We now proceed to discuss the symmetry-breaking pattern. To this aim, for a crystal in equilibrium, we fix the values of $\Phi_a$ and $\Psi$ as
\begin{equation} \label{eq:equilibrium_pm}
\Phi_a =  x_a\,, \quad \Psi  = t\,.
\end{equation} 
The equilibrium configuration \eqref{eq:equilibrium_pm} spontaneously breaks several of the symmetries, but preserves certain linear combinations of them. To be more precise, the symmetry breaking pattern can be characterized as follows:
\begin{equation}\begin{split}
\text{unbroken} &:  
     \begin{cases}
        \bar H = H +Q\,,  \quad  \bar J = J+L \,,\\
       \bar P_a = P_a + T_a -\frac{1}{2}\epsilon_{ab} x_b Q\,,\\
     \end{cases}\\
     \text{broken} &: 
     \begin{cases}
       Q\,,\quad  T_a \,,\quad L\,.
     \end{cases}
     \end{split}\label{eq:solidgenerators}
\end{equation}
To write down a nonlinear realization of this symmetry breaking pattern, we employ the coset construction method and parametrize the coset space as 
\begin{equation} \label{eq:parameter}
\Omega = e^{t \bar H} e^{ x^a \bar P_a} e^{ \phi Q} e^{ u_a  T_a} e^{ \theta L} \,.
\end{equation}
In the above parameterization, we have introduced four Goldstone fields: $\phi$, $u_a$, and $\theta$. After identifying symmetry-invariant fields by calculating the Maurer-Cartan form $\omega = \Omega^{-1} d \Omega$ (see \ref{app:coset} for details), we find that the effective field theory is
\begin{equation}\label{eq:effective}
\begin{split}
S&=-\frac{1}{2}\int d^2x dt \Big[ \partial_\mu\theta\partial^\mu\theta +    D_\mu u_iD^\mu u^{i} +  D_\mu\phi D^\mu\phi\Big]   \,, 
\end{split}
\end{equation}
where the covariant derivatives are
\begin{equation}
    D_\mu u_{i}=(\dot u_i,\partial_j u_i+\epsilon_{ji}\theta)\,, \quad D_\mu\phi=(\dot\phi,\partial_i\phi + \epsilon_{ij}u_j)\,, 
\end{equation} 
and the upper indices derivatives have been defined as follows: 
\begin{equation}\begin{split}
    \partial^\mu\theta&=(-f_\theta\dot\theta,m_\theta\partial_i\theta)\,, \quad D^\mu\phi=(-f_\phi\dot\phi,m_\phi D_i\phi)\,, \\
    D^\mu u^j&=(-f_u\dot u_j,C_{ijkl}D_k u_l)\,.
    \end{split}
    \label{eq:upper_indices}
\end{equation}
In the expressions above, $C_{ijkl}=C_1 \delta_{ij} \delta_{kl} + C_2 \delta_{i\langle k} \delta_{l \rangle j} + C_3 \epsilon_{ij} \epsilon_{kl}$ is the elastic tensor.

Notice that in the effective action \eqref{eq:effective}, $\epsilon_{ij}\theta$ and $\epsilon_{ij}u_j$ enter as Stueckelberg fields  for internal rotations and $Q$ transformations, respectively, signaling the presence of massive modes. In fact, integrating them out is equivalent to imposing the so-called inverse Higgs constraints associated with the symmetry-breaking pattern of the problem \cite{Ivanov1975,Low2002}
\begin{align}
\theta &= -\frac{1}{2} \epsilon_{ij} \partial_i u_j \,,\qquad 
u_i = \epsilon_{ij} \partial_j \phi\,.\label{eq:inv_higgs2}
\end{align}
The equation on the right of Eq.~\eqref{eq:inv_higgs2} implies that the displacements are transverse, $\partial_i u_i=0$, and therefore the density of the lattice sites is constant. This justifies the interpretation of Eq.~\eqref{eq:magneticAlegebra} as describing an incompressible solid.
The low-energy effective action takes the form of a Lifshitz theory 
\begin{equation}
    S_{\mathrm{gapless}} = \frac{1}{2}\int d^2xdt \Big[ f_\phi(\partial_0 \phi)^2 + \frac{1}{2}C_2 (\nabla^2 \phi)^2 \Big] \,.
\end{equation}
In fact, this theory has the same gapless spectrum ($\omega^2 \sim k^4$) as the quadrupole gauge theory discussed in the previous section.  In what follows, we will prove that not only does the gapless spectrum match, but also that the two theories are in fact dual.

\textit{Duality.}---We will now perform the dualization procedure for the action~\eqref{eq:effective}. 
The equations of motion read 
\begin{equation}
\label{eq:eomelasticity}
    \partial_\mu S^\mu   =\epsilon_{ij}\tau^{ij}\,,\quad
 \partial_\mu \tau^{\mu i}  =-\epsilon_{ij}K^j \,,\quad
 \partial_\mu K^\mu =0\,,
\end{equation}
where the crystal angular momentum, stress and charge currents are $S^\mu=\partial^\mu\theta$, $\tau^{\mu i}=D^\mu u^i$, $K^\mu=D^\mu\phi$ respectively, playing the role of Hubbard-Stratonovich fields in the dualization procedure. 
Moreover, we note that Eqs. \eqref{eq:eomelasticity} coincide with the Bianchi identities~\eqref{eq:bianchis} once we introduce the identification
\begin{equation}
\begin{split}
S^\mu &= \frac{1}{2}\epsilon^{\mu\nu\rho}F^{(0)}_{\nu\rho} \,, \quad \tau^{\mu i}= -\frac{1}{2}\epsilon^{\mu\nu\rho}\epsilon_{ij}F^{(1)}_{j\nu\rho}\,,\\
K^\mu &= \frac{1}{2}\epsilon^{\mu\nu\rho}F^{(2)}_{\nu\rho}\,.
\end{split}    
\end{equation}
Since the Bianchi identities of the quadrupole gauge theory coincide with the equations of motion \eqref{eq:eomelasticity}, we conclude that the dual description of the crystal will be in terms of gauge fields $A^{(0)},A_a^{(1)},A^{(2)}$ associated to $\theta,u_a,\phi$ respectively. Therefore, the incompressible solid phase is the elastic dual of the quadrupole gauge theory.

So far, we have considered regular crystal field configurations. However, in the presence of topological defects, $\theta$, $u_i$, and $\phi$ contain singular components that we denote $\theta^{(s)}$, $u_i^{(s)}$, and $\phi^{(s)}$. Implementing the dualization procedure (details of the derivation are presented in \ref{app:duality}) we arrive at the dual action
$S_{\mathrm{dual}}=S_{\mathrm{EM}}+S_{\mathrm{defects}}$, where $S_{\mathrm{EM}}$ takes the form of the quadrupole gauge theory~\eqref{eq:action} with $\lambda=0$ (albeit with a more general set of coefficients), while $S_{\mathrm{defects}}$ contains coupling between the gauge fields $A^{(0)}, A^{(1)}_a, A^{(2)}$ and the defects currents:  
\begin{equation} \label{eq:defect_currents}
\begin{split}
    J^{(0)\mu} & =-\epsilon^{\mu\nu\lambda}\partial_\nu\partial_\lambda \theta^{(s)}\,,\\
    J^{(1)\mu}_i &= \epsilon^{\mu\nu\lambda}\left(\epsilon_{li}\partial_\nu\partial_\lambda u_l^{(s)} - 2\delta_{i\lambda} \partial_\nu \theta^{(s)}\right)\,,\\
    J^{(2)\mu} &=-\epsilon^{\mu\nu\lambda}\left(\partial_\nu \partial_\lambda\phi^{(s)} + 2\delta_{\lambda k}\epsilon_{kj}\partial_\nu u^{(s)}_j\right)\,.
\end{split}
\end{equation}
In summary, the symmetric phase $(\lambda=0)$ of the action~\eqref{eq:action} describes an incompressible crystal, while the Higgs phase $(\lambda\neq0)$ corresponds to a situation in which the quadrupole defects form a condensate. Furthermore, in \ref{app:modes} we show that the low-energy action of the Higgs phase is equivalent to the traceless symmetric tensor gauge theory dualized by Pretko and Radzihovsky \cite{Pretko2018a}. Thus, ordinary elasticity theory emerges after Higgsing the quadrupole symmetry, which corresponds to a phase transition from an incompressible crystal to a compressible solid state hosting both longitudinal and transverse modes. 

Before closing this section, we briefly discuss the restrictions on the mobility of the topological defects in incompressible crystals. The conservation of monopole, dipole and quadrupole charges in the gauge theory is mapped onto the corresponding continuity equations for the topological defects:
\begin{equation}\label{eq:ward}
\partial_{\mu}J^{(0) \mu}=0\,, \quad \partial_{\mu} J^{(1)\mu}_a=J^{(0)}_a\,, \quad \partial_{\mu}J^{(2) \mu}=J^{(1)}_{aa}\,.
\end{equation}
The fractonic character of elastic defects can be explained as follows  \cite{Pretko2018a,Pretko2018b,Pretko2019,PhysRevLett.121.235301,PhysRevB.100.045119,Surowka2020,Nguyen_2020,Vijay}. The motion of monopoles (disclinations), $J^{(0)}_a\neq 0$, necessarily involves absorption or emission of dipoles (dislocations); similarly, the climb motion of dipoles, $J^{(1)}_{aa}\neq 0$, has to be accompanied by absorption or emission of quadrupoles (vacancies or interstitials). Therefore, in the symmetric phase, where all the defects are gapped, disclinations are immobile while dislocations can only move along their respective Burgers vector.

{\it Discussion.}---In this letter, inspired by fracton-elasticity duality \cite{Pretko2018a}, we have performed a detailed study of the gauge theory with monopole, dipole and quadrupole charges. We have demonstrated that the quadrupole gauge theory exhibits a peculiar anti-Higgs mechanism, in which the number of massless modes in the Higgs phase increases rather than decreases. Furthermore, we have proposed an elastic dual for the quadrupole gauge theory: an incompressible crystal phase, in which lattice translations obey the magnetic algebra \eqref{eq:magneticAlegebra}. Finally, we showed that the Higgs phase of the quadrupole gauge theory is equivalent to the symmetric tensor gauge theory that is known to be dual to ordinary elasticity \cite{Pretko2018a}. Importantly, our construction naturally incorporates the quadrupole symmetry in the dual gauge theory of elasticity, which is required in order to conform with the phenomenology of topological defects \cite{Pretko2018a} (see also \cite{PhysRevLett.121.235301,PhysRevB.100.045119} for different approaches to the problem).

We now highlight some promising directions for future research: first, it would be interesting to determine the generic property of a gauge group that gives rise to the anti-Higgs mechanism; and second, to identify systems exhibiting it. 

In fact, we have identified two systems that appear to fall within our classification of incompressible crystals. One example of such a system is given by the Wigner crystal of charged particles coupled to a dynamical $U(1)$ gauge field, which is a $2$--dimensional version of the theory described in \cite{Du2024} -- in this case, $Q$ can be identified as the magnetic $0$-form symmetry. It would be interesting to explore the applicability of the anti-Higgs mechanism to the understanding of phase transitions in Wigner crystals. In particular, \cite{Kivelson2024} shows that interstitials can proliferate in a two-dimensional Wigner crystal above a critical electron density. Since the physical systems exist in three spatial dimensions, it will also be useful to generalize our classification of incompressible solids to three-dimensional systems.

Another example of an incompressible crystal is given by the emergent vortex crystals formed in rotating superfluids where charge $Q$ is the global $U(1)$ bosonic symmetry \cite{10.21468/SciPostPhys.5.4.039,Nguyen_2020,Radzihovsky_2024}. However, vortex crystals break time-reversal symmetry, leading to a different structure of the effective theory and ultimately resulting in a distinct dual gauge theory where the anti-Higgs mechanism is absent \cite{Nguyen_2020}.

\acknowledgments 

The authors thank Sergej Moroz and Piotr Surówka for discussions. A.G. has been supported through a stipend from the International Max Planck Research School (IMPRS) for Quantum Dynamics and Control hosted at the Max Planck Institute for the Physics of Complex Systems. P.M. and L.T. have been supported in part by the Deutsche Forschungsgemeinschaft through the cluster of excellence ct.qmat (Exzellenzcluster 2147, Project No. 390858490). F.P.-B. has received funding from the Norwegian Financial Mechanism 2014-2021 via the NCN, POLS Grant 2020/37/K/ST3/03390.

\bibliographystyle{biblev1}
\bibliography{quadrupole}

%================================================================================
\newpage
\phantom{a}
\newpage
%%%%%%%%%%%%%%%%%%%%%%%%%%%%%%%%%%%%%%%%
\setcounter{figure}{0}
\setcounter{equation}{0}

\renewcommand{\thetable}{S\arabic{table}}
\renewcommand{\thefigure}{S\arabic{figure}}
\renewcommand{\theequation}{S\arabic{equation}}
\renewcommand{\thepage}{S\arabic{page}}

\renewcommand{\thesection}{S\arabic{section}}

\onecolumngrid

\begin{center}
{\large \bf Supplemental Material:\\
Quadrupole gauge theory: anti-Higgs mechanism and elastic dual}\\

\vspace{0.3cm}

\setcounter{page}{1}
\end{center}

\label{pagesupp}
\section{Gauging the quadrupole algebra with the coset construction method}
\label{app:coset_quadrupole}
In this supplementary material, we gauge the quadrupole algebra using the coset construction formalism. We begin by extending the symmetry group to include rotations with generators $L_{ab}$, such that the complete set of nontrivial commutators is given by:
\begin{equation}
\begin{split}
[P_a,Q^{(1)}_b]&=\delta_{ab}Q^{(0)}\,, \hspace{2.8cm} [P_a,Q^{(2)}]=Q^{(1)}_a\,, \hspace{1.5cm} [L_{ab},P_c]=\delta_{ac}P_b-\delta_{bc} P_a\,, \\
[Q^{(1)}_a, L_{bc}]&=\delta_{ac}Q^{(1)}_b-\delta_{ab}Q^{(1)}_c\,, \hspace{1.25cm} [L_{ab},L_{cd}]=\delta_{ac}L_{bd}-\delta_{ad}L_{bc}-\delta_{bc}L_{ad}+\delta_{bd}L_{ac}\,.
\end{split}
\end{equation}
Let us denote by $G$ the full symmetry group and by $H$ its subgroup generated by $L_{ab}$. Following the standard procedure for gauging spacetime symmetries \cite{Ivanov1975}, we consider the coset $G/H$ parametrized as
\begin{equation}\label{eq:coset}
\Omega=e^{t H}e^{x_aP_a}e^{\psi^{(0)} Q^{(0)}}e^{\psi^{(1)}_a Q^{(1)}_a}e^{\psi^{(2)} Q^{(2)}}.
\end{equation}
The fields $\psi^{(0)}$, $\psi^{(1)}_a$, $\psi^{(2)}$ are understood as the Stueckelberg fields of the broken $Q^{(0)}$, $Q^{(1)}_a$, and $Q^{(2)}$ symmetries, respectively. In order to construct the invariant building blocks of the theory, we consider the Maurer-Cartan 1-form for the local symmetry
\begin{equation}
     \omega =\Omega^{-1} (d+\mathcal{A}) \Omega\,,
\end{equation} 
where we have introduced a connection $\mathcal{A}$ given in terms of the theory's generators as
\begin{equation}
\mathcal{A}=\Tilde{A}^{(0)}Q^{(0)}+\tilde{A}^{(1)}_aQ^{(1)}_a +A^{(2)} Q^{(2)}\,,
\end{equation}
and set $\tilde{A}^{(0)}=A^{(0)}+x_aA^{(1)}_a+\frac12 x^2 A^{(2)}$ and $\tilde{A}^{(1)}_a=A^{(1)}_a+x_aA^{(2)}$ for our convenience. We call $A^{(0)}_{\mu}$, $A^{(1)}_{a\mu}$, and $A^{(2)}_\mu$ the monopole, dipole, and quadrupole gauge fields, respectively.

The Maurer-Cartan form $\omega$ is gauge-invariant by construction and its expansion in the basis of the operators is
\begin{equation}
\omega=dt H+dx_a P_a+VQ^{(0)}+W_aQ^{(1)}_a+UQ^{(2)}\,.
\end{equation}
Making use of the commutations relations and the Baker-Campbell-Hausdorff formula, we obtain
\begin{equation}
\begin{split}\label{eq:stuckelberg}
     V = d\psi^{(0)} + dx_a \psi^{(1)}_a + A^{(0)}\,, \quad  W_a = d\psi^{(1)}_a + dx_a\psi^{(2)}  + A^{(1)}_{a}\,, \quad U = d\psi^{(2)}+A^{(2)}\,.
    \end{split}
\end{equation}
The field strengths can be identified as the $Q^{(0)}$, $Q^{(1)}_a$, and $Q^{(2)}$ components of the curvature 2-form $\mathcal F = d\omega+\frac{1}{2}[\omega,\omega]$, which are given by
\begin{equation}
\begin{split}
F^{(0)}&=dV + dx_a\wedge W_a=dA^{(0)} + dx_a\wedge A^{(1)}_a\,,\\
F^{(1)}_{a}&=d W_a + dx_a \wedge U=d A^{(1)}_a + dx_a \wedge A^{(2)}\,,\\
F^{(2)}&=dU=dA^{(2)}\,.
\label{app:field_strengths}
\end{split}    
\end{equation}
Importantly, the field strengths do not depend on the Stuckelberg fields $\psi^{(0)}$, $\psi^{(1)}_a$, $\psi^{(2)}$, so they are gauge-invariant independently of whether some of the symmetries are broken or not. Together, Eqs. \eqref{eq:stuckelberg} and \eqref{app:field_strengths} form the complete set of gauge-invariant quantities from which the action is constructed.

Let us now examine the transformation properties of the Stueckelberg and gauge fields under an infinitesimal gauge transformation:
\begin{equation}\label{eq:infinitesimalGauge}
g\approx1+\tilde{\lambda}^{(0)} Q^{(0)}+\tilde{\lambda}^{(1)}_aQ^{(1)}_a+\lambda^{(2)} Q^{(2)},   
\end{equation}
where we have parametrized $\tilde{\lambda}^{(1)}_a=\lambda^{(1)}_a +x_a\lambda^{(2)}$ and $\tilde{\lambda}^{(0)}=\lambda^{(0)}+x_a\lambda^{(1)}_a+\frac12 x^2 \lambda^{(2)}$. Under the action of \eqref{eq:infinitesimalGauge} the coset element \eqref{eq:coset} transforms as $\Omega \rightarrow g \Omega $. After calculating $g \Omega$ with the help of the Baker-Campbell-Hausdorff formula, we find
\begin{equation}
\delta \psi^{(0)} = \lambda^{(0)}, \hspace{2cm} \delta \psi^{(1)}_a = \lambda^{(1)}_a, \hspace{2cm} \delta \psi^{(2)} = \lambda^{(2)}.
\end{equation}
On the other hand, the connection transforms as $\mathcal A \rightarrow g \mathcal A g^{-1}+g dg^{-1}$. Applying the formula to the infinitesimal gauge transformation \eqref{eq:infinitesimalGauge} we obtain the following transformation properties for the gauge fields
\begin{equation}
\delta A^{(0)} =  -d\lambda^{(0)}-dx_a\lambda^{(1)}_a\,, \hspace{2cm} \delta A^{(1)}_a  = -d\lambda^{(1)}_a - dx_a\lambda^{(2)} \,, \hspace{2cm} \delta A^{(2)} = -d\lambda^{(2)}\,.
\end{equation}
Finally, in the main text we have assumed the absence of the monopole and dipole Stueckelberg fields, which is equivalent to setting $\psi^{(0)}=0$ and $\psi^{(1)}_a=0$ in the coset parameterization Eq. \eqref{eq:coset}.

%%%%%%%%%%%%%%%%%%%%%%%%%%%%%%%%%%%%%%%%
\section{Low-energy effective action of quadrupole gauge theory}
\label{app:modes}
In this section, we will find the gapless modes by the means of integrating out the massive degrees of freedom in the action~(\ref{eq:action}). We will use round brackets to denote symmetrization: $A_{(ab)}=\frac12\left(A_{ab}+A_{ba}\right)$, square brackets to denote antisymmetrization: $A_{[ab]}=\frac12\left(A_{ab}-A_{ba}\right)$, and angle brackets to denote the traceless symmetric part: $A_{\langle ab\rangle}=A_{(ab)}-\frac{1}{d}\delta_{ab}A_{ii}$, where $d$ is the spatial dimension and $A_{ii}$ is the trace of $A_{ab}$.

We start by writing the action as
\begin{equation}
   S = \frac12\int d^d x\left[l^{-2}E^{(0)}_a E^{(0)}_a-\frac{l^{-2}}{2} F^{(0)}_{ab} F^{(0)}_{ab}+E^{(1)}_{ab}E^{(1)}_{ab}-\frac{1}{2} F^{(1)}_{abc}F^{(1)}_{abc}+L^2E^{(2)}_aE^{(2)}_a - \frac{L^2}{2} F^{(2)}_{ab} F^{(2)}_{ab}+\lambda^2 U_0^2-\lambda^2 U_a^2\right]\,,
   \label{eq:effective_action}
\end{equation}
where
\begin{equation}
    E^{(0)}_a \equiv F^{(0)}_{a0}\,,~~~~E^{(1)}_{ab} \equiv F^{(1)}_{ab0}\,,~~~~E^{(2)}_a \equiv F^{(2)}_{a0}\,.
\end{equation}
The dipole terms can be further decomposed in the following way. For the electric field:
\begin{equation}
    \frac12\left( E^{(1)}_{ab} \right)^2 = \frac12 \left(E^{(1)}_{\langle ab\rangle}+ E^{(1)}_{[ ab]}+\frac{\delta_{ab}}{d}E^{(1)}_{ii}\right)^2= \frac12 \left( E^{(1)}_{\langle ab \rangle} \right)^2+\frac12 \left( E^{(1)}_{[ab]} \right)^2+\frac{1}{2d} \left( E^{(1)}_{ii} \right)^2 \,,
\end{equation}
where by a square we mean a sum over the indices, and where
\begin{equation}
\begin{split}
    E^{(1)}_{\langle ab\rangle} = \partial_{\langle b} A^{(1)}_{a\rangle0}-\partial_0A^{(1)}_{\langle ab\rangle}\,, \qquad   E^{(1)}_{[ ab]} = \partial_{[ b} A^{(1)}_{a]0}-\partial_0A^{(1)}_{[ ab]}\,, \qquad
    E^{(1)}_{ii} = d A^{(2)}_0+\partial_iA^{(1)}_{i0}-\partial_0A^{(1)}_{ii}\,.
\end{split}
\end{equation}
A somewhat similar decomposition of $\left( F^{(1)}_{abc}\right)^2$ can also be performed:
\begin{equation}
    \left( F^{(1)}_{abc}\right)^2 = \left(F^{(1)\mathrm{S}}_{abc}\right)^2+ \frac13\left(F^{(1)\mathrm{A}}_{abc}\right)^2+\frac{2}{d-1}\left(F^{(1)}_{iia}\right)^2 \,,
    \label{eq:Fabc_square}
\end{equation}
where
\begin{equation}
\begin{split}
    F^{(1)\mathrm{S}}_{abc} &\equiv \left(\frac12F^{(1)}_{abc}+\frac12F^{(1)}_{bac}+\frac12 F^{(1)}_{cba}-\frac{\delta_{ab}}{d-1}F^{(1)}_{iic}+\frac{\delta_{ac}}{d-1}F^{(1)}_{iib}\right)\\
    &=\left(\partial_bA^{(1)}_{\langle ac\rangle}-\partial_cA^{(1)}_{\langle ab\rangle}+\partial_aA^{(1)}_{[bc]}\right)-\frac{\delta_{ab}}{d-1}\left(\partial_iA^{(1)}_{\langle ic\rangle}+\partial_iA^{(1)}_{[ic]}\right)+\frac{\delta_{ac}}{d-1}\left(\partial_iA^{(1)}_{\langle ib\rangle}+\partial_iA^{(1)}_{[ib]}\right)\,,\\
    F^{(1)\mathrm{A}}_{abc} & \equiv\left(\frac12F^{(1)}_{abc}-\frac12F^{(1)}_{bac}-\frac12 F^{(1)}_{cba}\right) = \partial_b A^{(1)}_{[ac]}-\partial_cA^{(1)}_{[ab]}-\partial_aA^{(1)}_{[bc]}\,, \\
    F^{(1)\mathrm{A}}_{iia} & = \partial_i A^{(1)}_{\langle ia \rangle}+\partial_i A^{(1)}_{[ ia] }+(d-1)\left(A^{(2)}_a-\frac{1}{d}\partial_a A^{(1)}_{ii}\right)\,.
\end{split}
\end{equation}
This decomposition is done in such a way that $F^{(1)\mathrm{S}}_{abc}$ and $F^{(1)\mathrm{A}}_{abc}$ are both antisymmetric in the last two indices and traceless in the first two indices $F^{(1)\mathrm{S}}_{iic}=F^{(1)\mathrm{A}}_{iic}=0$, while $F^{(1)\mathrm{A}}_{abc}$ is additionally antisymmetric in the first two indices $F^{(1)\mathrm{A}}_{abc}=-F^{(1)\mathrm{A}}_{bac}$. With these decompositions, the action is expressed in a form convenient for integrating out massive degrees of freedom. 

Let us first discuss the unbroken theory $\lambda=0$. By way of example, let us show how $A^{(2)}_0$ can be integrated out. The partition function reads
\begin{equation}
    \mtc Z = \int [\mathcal{D} A^{(0)}_\mu \mathcal{D} A^{(1)}_{a\mu} \mathcal{D} A^{(2)}_\mu] \exp(iS)\,.
\end{equation}
The square brackets around the volume element denote integrating over equivalence classes of field configurations that are related by a gauge transformation. Now, notice that 
\begin{equation}
    A^{(2)}_0 = \frac{1}{d}E^{(1)}_{ii}-\frac{1}{d}\partial_iA^{(1)}_{i0}+\frac{1}{d}\partial_0A^{(1)}_{ii}\,.
    \label{eq:change_of_vars-action}
\end{equation}
Performing the change of variables $A^{(2)}_0\rightarrow E^{(1)}_{ii}$ in the path integral does not change the measure except for the unimportant prefactor, $\mathcal{D} A^{(2)}_0 \rightarrow \frac{1}{d}\mathcal{D} E^{(1)}_{ii}$, because the remaining terms in Eq.~(\ref{eq:change_of_vars-action}) do not contribute to the Jacobian of the transformation. Moreover, $\frac{1}{d}E^{(1)}_{ii}$ contains only the gauge-invariant part of $A^{(2)}_0$, so by performing this change of variables we avoid the complications related to evaluating the path integral over the gauge-dependent quantity $A^{(2)}_0$. Integrating over $E^{(1)}_{ii}$ and ignoring higher-order terms in the derivative expansion produces an effective action, which turns out to be equal to the action~(\ref{eq:effective_action}) with the substitution $A^{(2)}_0\rightarrow -\frac{1}{d}\partial_iA^{(1)}_{i0}+\frac{1}{d}\partial_0A^{(1)}_{ii}$. 

The abovementioned procedure can be used to integrate out $A^{(1)}_{a0}$, $A^{(1)}_{[ab]}$, $A^{(2)}_0$ and $A^{(2)}_a$. Interestingly, this procedure not only eliminates $E^{(0)}_a$, $F^{(0)}_{ab}$, $E^{(1)}_{ii}$ and $F^{(1)}_{iia}$ from the action, but also sets the antisymmetric dipolar terms $E^{(1)}_{[ab]}$ and $F^{(1)\mathrm{A}}_{abc}$ to zero and eliminates the gauge field $A^{(1)}_{ii}$ from the action. The reason for the disappearance of $A^{(1)}_{ii}$ is that the action has to remain quadrupole-gauge invariant after eliminating $A^{(2)}_\mu$, as discussed also in the main text. In the end,
\begin{equation}
    S_{\mathrm{eff}}^{(\lambda=0)} = \frac12\int d^d x\left[E^{(1)}_{\langle ab\rangle}E^{(1)}_{\langle ab\rangle}-\frac{1}{2}F^{(1)\mathrm{S}}_{abc}F^{(1)\mathrm{S}}_{abc}+L^2E^{(2)}_aE^{(2)}_a - \frac{L^2}{2} F^{(2)}_{ab} F^{(2)}_{ab}\right]\,,
\end{equation}
where all the fields are now expressed through $A^{(0)}_0$, $A^{(0)}_a$ and $A^{(1)}_{\langle ab\rangle}$ only. At this point, a lengthy but straightforward calculation shows that $E^{(2)}_a=-\frac{1}{d-1}\partial_iE^{(1)}_{\langle ia\rangle}$ and $F^{(2)}_{ab}=-\frac{1}{d-2}\partial_iF^{(1)\mathrm{S}}_{iab}$ for $d> 2$, so in the long-wavelength limit the terms $E^{(2)}_aE^{(2)}_a - \frac12 F^{(2)}_{ab} F^{(2)}_{ab}$ are subleading and can be neglected (if $d> 2$). 

Let us first consider $d>2$. Then, the low-energy action is
\begin{equation}
    S_{\mathrm{eff}}^{(\lambda=0,\,d>2)} = \frac{1}{2}\int d^d x\left[E^{(1)}_{\langle ab\rangle}E^{(1)}_{\langle ab\rangle}-\frac{1}{2}F^{(1)\mathrm{S}}_{abc}F^{(1)\mathrm{S}}_{abc}\right]\,.
\end{equation}
Varying the action with respect to $A^{(0)}_0$ and $A^{(1)}_{\langle ab\rangle}$ produces the Gauss's law $\partial_a\partial_b E^{(1)}_{\langle ab\rangle}=0$ and the Amp\`{e}re's law $\partial_0E^{(1)}_{\langle ab\rangle}-\partial_c F^{(1)}_{\langle ab\rangle c}=0$, respectively. If we fix the gauge by setting $A^{(0)}_a=A^{(0)}_0=0$, all the field strengths can be expressed through $A^{(1)}_{\langle ab\rangle}$ alone; namely,
\begin{equation}
\begin{split}
    E^{(1)}_{\langle ab\rangle} =-\partial_0 A^{(1)}_{\langle ab\rangle}\,, \qquad
    F^{(1)S}_{abc} = \left(\partial_bA^{(1)}_{\langle ac\rangle}-\partial_cA^{(1)}_{\langle ab\rangle}\right)-\frac{1}{d-1}\left(\delta_{ab}\partial_iA^{(1)}_{\langle ic\rangle}-\delta_{ac}\partial_iA^{(1)}_{\langle ib\rangle}\right)\,.
\end{split} \label{eq:app_field_strengths}
\end{equation}
Thus, in $d=3$ there are $5$ degrees of freedom corresponding to $A^{(1)}_{\langle ab\rangle}$, obeying one Gauss's law constraint, for the total of $4$ modes. Assuming a perturbation propagating with frequency $\omega$ and wavevector $\mathbf k$ along the $x$ direction, it is straightforward to calculate the dispersion relation of the modes:
\begin{center}
\begin{tabular}{ |c|c| } 
\hline
 Mode & Dispersion relation \\
 \hline
 $A^{(1)}_{\langle xy\rangle}$, $A^{(1)}_{\langle xz\rangle}$ & $\omega^2 = \frac14 |\mathbf{k}|^2$ \\
 \hline
 $A^{(1)}_{\langle yy\rangle}$, $A^{(1)}_{\langle yz\rangle}$ & $\omega^2 = |\mathbf{k}|^2$ \\
 \hline
\end{tabular}
\end{center}

In the case $d=2$, $F^{(1)\mathrm{S}}_{abc}$ is identically zero. The reason for this is the asymmetry in the last two indices, which means that $F^{(1)\mathrm{S}}_{xxy}$ and $F^{(1)\mathrm{S}}_{yyx}$ are the only non-trivial components, which in combination with the tracelessness condition enforces $F^{(1)\mathrm{S}}_{xxy}+F^{(1)\mathrm{S}}_{yyy}=F^{(1)\mathrm{S}}_{xxy}=0$, and analogically, $F^{(1)\mathrm{S}}_{yyx}+F^{(1)\mathrm{S}}_{xxx}=F^{(1)\mathrm{S}}_{yyx}=0$. Therefore, for $d=2$, $F^{(2)}_{ab} F^{(2)}_{ab}$ is a leading-order term in the low-energy action,
\begin{equation}
    S_{\mathrm{eff}}^{(\lambda=0,\,d=2)}  = \frac12\int d^d x\left[E^{(1)}_{\langle ab\rangle}E^{(1)}_{\langle ab\rangle}-\frac{L^2}{2} F^{(2)}_{ab} F^{(2)}_{ab}\right]\,.
\end{equation}
Amp\`{e}re's law now takes the form $\partial_0E^{(1)}_{\langle ab\rangle}-L^2\partial_c \partial_{\langle a}F^{(2)}_{b\rangle c}=0$, where, in the gauge $A^{(0)}_\mu=0$,
\begin{equation}
    F^{(2)}_{ab} = -\partial_c \partial_a A^{(1)}_{\langle cb \rangle}+\partial_c \partial_b A^{(1)}_{\langle ca \rangle}\,.
\end{equation}
Calculating the modes as before, we now obtain a quadratic dispersion relation
\begin{center}
\begin{tabular}{ |c|c| } 
\hline
 Mode & Dispersion relation \\
 \hline
 $A^{(1)}_{\langle xy\rangle}$ & $\omega^2 = \frac{1}{2}L^2 |\mathbf{k}|^4$\\
 \hline
\end{tabular}
\end{center}

Now we turn our attention to the Higgs phase, which means that the terms $\frac{1}{2}\lambda^2\left(A^{(2)}_0+\partial_0\psi^{(2)}\right)^2-\frac{1}{2}\lambda^2\left(A^{(2)}_a+\partial_a\psi^{(2)}\right)^2$ appear in the action. The presence of these terms modifies the procedure, via which the field $A^{(2)}_\mu$ is integrated out. Focusing on $A^{(2)}_0$, we notice the existence of terms $\frac{1}{2}m_1\left(A^{(2)}_0-X\right)^2+\frac{1}{2}m_2\left(A^{(2)}_0-Y\right)^2$, where $m_1=d$, $m_2=\lambda^2$, $X=-\frac{1}{d}\partial_iA^{(1)}_{i0}+\frac{1}{d}\partial_0A^{(1)}_{ii}$, $Y=-\partial_0\psi^{(2)}$, in the action. We will make use of the following identity:
\begin{equation}
    \frac{m_1}{2}\left(A^{(2)}_0-X\right)^2+\frac{m_2}{2}\left(A^{(2)}_0-Y\right)^2 = \frac{m_1+m_2}{2}\left(A^{(2)}_0-\frac{m_1 X+m_2 Y}{m_1+m_2}\right)^2 + \frac{m_1m_2}{2(m_1+m_2)}\left(X-Y\right)^2\,.
\end{equation}
Integrating out the massive mode amounts to replacing the two mass terms with one term proportional to $\left(X-Y\right)^2$, while setting $A^{(2)}_0 \rightarrow \frac{m_1 X+m_2 Y}{m_1+m_2}$ in the remaining terms. The same procedure can be applied to $A^{(2)}_{a}$. In the end, the low-energy action is
\begin{equation}
    S_{\mathrm{eff}}^{(\lambda\neq0)}  = \frac{1}{2}\int d^d x\left[E^{(1)}_{\langle ab\rangle}E^{(1)}_{\langle ab\rangle}-\frac{1}{2}F^{(1)\mathrm{S}}_{abc}F^{(1)\mathrm{S}}_{abc}+\frac{d+\lambda^2}{d\lambda^2}\tilde{E}^{(1)}\tilde{E}^{(1)} - \frac{(d-1)+\lambda^2}{(d-1)\lambda^2}\tilde{B}^{(1)}_a\tilde{B}^{(1)}_a\right]\,,
    \label{eq:app_action_broken}
\end{equation}
where 
\begin{equation}
\begin{split}
    \tilde{E}^{(1)} & = \frac{d\lambda^2}{d+\lambda^2}\left(-\partial_0\psi^{(2)}+\frac{1}{d}\partial_0\partial_iA^{(0)}_i-\frac{1}{d}\partial_i^2A^{(0)}_0-\frac{1}{d}\partial_0A^{(1)}_{ii}\right)\,,\\
    \tilde{B}^{(1)}_a & = -\epsilon_{ab}\frac{(d-1)\lambda^2}{(d-1)+\lambda^2}\left(-\partial_b \psi^{(2)} -\frac{1}{2(d-1)}\left(\partial_i^2A^{(0)}_{b}-\partial_b\partial_iA^{(0)}_i\right)+\frac{1}{d-1}\partial_iA^{(1)}_{\langle i b\rangle}-\frac{1}{d}\partial_bA^{(1)}_{ii}\right)\,.
    \label{eq:EH_tilde}
\end{split}
\end{equation}
In comparison with the symmetric case, there are two more terms in the action and two more degrees of freedom, $\psi$ and $A^{(1)}_{ii}$. To calculate the dispersion relation, it is then convenient to fix the gauge $A_a^{(0)}=A_0^{(0)}=\psi^{(2)}=0$, so that all the fields can be expressed in terms of $A^{(1)}_{\langle ab\rangle}$ and $A^{(1)}_{ii}$. In three dimensions we obtain
\begin{center}
\begin{tabular}{ |c|c| } 
\hline
 Mode & Dispersion relation \\
 \hline
 $A^{(1)}_{ii}$ & $\omega^2 = |\mathbf{k}|^2$ \\
 \hline
 $A^{(1)}_{\langle xy\rangle}$, $A^{(1)}_{\langle xz\rangle}$ & $\omega^2 = \frac{1+\lambda^2}{4+2\lambda^2} |\mathbf{k}|^2$ \\
 \hline
 $A^{(1)}_{\langle yy\rangle}$, $A^{(1)}_{\langle yz\rangle}$ & $\omega^2 = |\mathbf{k}|^2$ \\
 \hline
\end{tabular}
\end{center}
and in two dimensions we obtain
\begin{center}
\begin{tabular}{ |c|c| } 
\hline
 Mode & Dispersion relation \\
 \hline
 $A^{(1)}_{ii}$ & $\omega^2 = |\mathbf{k}|^2$ \\
 \hline
 $A^{(1)}_{\langle xy\rangle}$ & $\omega^2 = \frac{\lambda^2}{2+2\lambda^2} |\mathbf{k}|^2$ \\
 \hline
\end{tabular}
\end{center}

In the closing paragraph of this section, we explain the connection between the action~(\ref{eq:app_action_broken}) and the rank-2 tensor gauge theory which was shown by Pretko and Radzihovsky in Ref.~\cite{Pretko2018a} to be dual to ordinary two-dimensional elasticity. Varying the action~(\ref{eq:app_action_broken}) with respect to $A^{(0)}_0$ and $A^{(1)}_{\langle ab\rangle}$ produces the Gauss's law and the Amp\`{e}re's law, respectively:
\begin{equation}
    \partial_a\partial_b\left(E^{(1)}_{\langle ab\rangle}+\frac{\delta_{ab}}{d}\tilde{E}^{(1)}\right)=0\,,~~~~
    \partial_0E^{(1)}_{\langle ab\rangle}-\partial_c F^{(1)\mathrm{S}}_{\langle ab\rangle c}-\frac{1}{d-1}\epsilon_{c\langle a}\partial_{b\rangle}\tilde{B}^{(1)}_{c}=0\,.
    \label{eq:app_maxwell_1}
\end{equation}
Varying the action with respect to $A^{(1)}_{ii}$ gives the quadrupole conservation equation
\begin{equation}
    \partial_0\tilde{E}^{(1)}-\epsilon_{ab}\partial_a\tilde{B}^{(1)}_b=0\,.
    \label{eq:app_maxwell_2}
\end{equation}
Two more equations of motion can be obtained by varying the action with respect to $A^{(0)}_a$ and $\psi^{(2)}$, but as it turns out, they can both be derived from Eqs.~(\ref{eq:app_maxwell_1}) and~(\ref{eq:app_maxwell_2}) and are in this sense redundant. For this reason, an action that leads to the same dynamics can be written after gauge-fixing $A^{(0)}_a=0$ and $\psi^{(2)}=0$, while retaining $A^{(0)}_0$ and $A^{(1)}_{(ab)}$. The requirement $A^{(0)}_a=0$ and $\psi=0$ leaves us with a residual gauge freedom: 
\begin{equation}
    \delta A^{(0)}_{0}=-\partial_0 \lambda^{(0)}\,,~~~~\delta A^{(1)}_{(ab)} = \partial_a\partial_b \lambda^{(0)}\,.
    \label{eq:app_pretko_cite1}
\end{equation}
In the end, a low-energy action equivalent to~(\ref{eq:app_action_broken}) in two dimensions reads
\begin{equation}
    S_{\mathrm{eff,alternative}}^{(\lambda\neq0)}  = \frac{1}{2}\int d^d x\left[E_{\langle ab\rangle}E_{\langle ab\rangle}+\frac{\lambda^2}{2(2+\lambda^2)}E_{ii}E_{ii} - \frac{\lambda^2}{1+\lambda^2}B_aB_a\right]\,,
    \label{eq:app_pretko_cite2}
\end{equation}
where 
\begin{equation}
    E_{ab} = -\partial_i^2A^{(0)}_0-\partial_0A^{(1)}_{(ab)}\,,~~~~
    B_a  = \epsilon_{bc}\partial_b A^{(1)}_{(ca)}\,.
    \label{eq:app_pretko_cite3}
\end{equation}
Equations~(\ref{eq:app_pretko_cite1})-(\ref{eq:app_pretko_cite3}) express the symmetric rank-2 tensor gauge theory of Ref.~\cite{Pretko2018a}.

%%%%%%%%%%%%%%%%%%%%%%%%%%%%%%%%%%%%%%%%
\section{Maurer-Cartan form and effective action for incompressible solids}\label{app:coset}
In order to identify the invariant building blocks that one can use to construct the effective action, we introduce the Maurer-Cartan form $\omega = \Omega^{-1} d \Omega$, where $\Omega = e^{t \bar H} e^{ x_a \bar P_a} e^{ \phi Q} e^{ u_a  T_a} e^{ \theta L}$ is the coset representative. The Maurer-Cartan form takes values in the Lie-algebra and can therefore be expanded in the basis of generators,
\begin{equation}
    \omega = dt \bar H + dx_a \bar P_a   + \omega^Q Q + \omega^T_a T_a+ \omega^L L\,. 
\end{equation}
The components associated with the broken generators are related to the covariant derivatives of the Goldstone fields,
\begin{equation}
    \omega^Q =  (D_0 \phi) dt + (D_a \phi) dx_a\,, \quad \omega^T_a = (D_0 u_a) dt + (D_b u_a)   dx_b  \,, \quad \omega^L = (D_0 \theta ) dt + (D_a \theta)   dx_a \,.
\end{equation}
After evaluating the Maurer-Cartan form explicitly we can read off the covariant derivatives and construct an effective action that is exactly invariant under the whole symmetry group. Below we summarize the nontrivial commutation relations between the generators,
\begin{equation} \begin{gathered}
 \label{eq:commutation}
[T_a\,, T_b] =    \epsilon_{ab} Q\,, \quad [ T_a\,, \bar P_b]=    \epsilon_{ab} Q\,, \\
[L\,, T_a] = \epsilon_{ab} T_b\,, \quad [L,\bar P_a]  = \epsilon_{ab} T_b\,, \quad
[\bar J, \bar P_a] = \epsilon_{ab} \bar P_b \,.
\end{gathered}
\end{equation} 
From the commutation relations \eqref{eq:commutation} we derive the following identities,
\begin{equation}\begin{split}
    e^{-u_a T_a} d e^{u_a T_a} &= d u_a T_a + \frac{1}{2}  \epsilon_{ab}  u_b du_a Q \,, \\
    e^{-u_b T_b} \bar P_a e^{u_b T_b} &= \bar P_a + \epsilon_{ab} u_b Q\,, \\
    e^{-\theta L}\bar P_a  e^{\theta L} &= \bar P_a + (\theta_{ab} - \delta_{ab})T_b\,,
    \end{split}
\end{equation}
where $\theta_{ab} \in SO(2)$ is a rotation matrix $\theta_{ab} = (e^{\theta L})_{ab}=\delta_{ab} + \theta \epsilon_{ab} + \mathcal{O}(\theta^2)$. 

With these relations at hand it is straightforward to compute the Maurer-Cartan form. After doing so, we find the following expressions for the components: 
\begin{equation}
    \begin{split}
        \omega^Q &= d \phi +  \epsilon_{ab} u_b dx_a + \frac{1}{2}  \epsilon_{ab}  u_b du_a \,, \\
        \omega^T_a &= d(u_b+x_b) \theta_{ab} - dx_a\,, \\ 
        \omega^L &= d \theta \,.
    \end{split}
\end{equation}
Finally, we can read off the covariant derivatives of the Goldstone fields as
\begin{equation} 
    \begin{split}
        D_0 \phi &= \partial_0 \phi + \frac{1}{2} \epsilon_{ab}  u_b \partial_0 u_a  \approx \partial_0 \phi\,, \\
        D_a \phi &= \partial_a \phi +  \epsilon_{ab} u_b + \frac{1}{2}  \epsilon_{bc}  u_c \partial_a u_b  \approx \partial_a \phi  +  \epsilon_{ab} u_b\,, \\
        D_0 u_a &=\partial_0 u_b \theta_{ab}  \approx \partial_0 u_a\,, \\ 
        D_a u_b &= \partial_a u_c \theta_{bc} + \theta_{ab} -\delta_{ab} \approx \partial_a u_b + \epsilon_{ab} \theta \,, \\
       D_0 \theta &= \partial_0 \theta \approx \partial_0 \theta \,, \\ 
        D_a \theta &= \partial_a \theta \approx \partial_a \theta\,.
    \end{split}
\end{equation}
The symbol “$\approx$” denotes the linearization of the derivatives. The effective action is constructed by taking inner products of the covariant derivatives while also preserving the parity $\big(P:x \longleftrightarrow y\big)$ and time reversal $\big(T:t \longrightarrow-t\big)$ symmetries. In particular, a quadratic action is the one proposed in the main text, 
\begin{equation}\label{eq:effective2}
S=\frac{1}{2}\int d^2xdt\Big[f_{\theta}(\partial_0 \theta)^2+f_u(\partial_0 u_i)^2+f_\phi(\partial_0 \phi)^2-m_{\theta}(\partial_i\theta)^2-C_{ijkl}D_i u_{j}D_k u_{l}-m_{\phi}(D_i\phi)^2 \Big] \,. 
\end{equation}

\section{Gauge-elasticity duality with defects}\label{app:duality}

Legendre-transforming the action~(\ref{eq:effective2}), we obtain the Hubbard-Stratonovich action
\begin{equation}
\begin{split}
S_{\mathrm{HS}}=\frac{1}{2}\int d^2xdt&\left[\frac{1}{m_{\theta}}\left(S^i\right)^2+C^{-1}_{ijkl}\tau^{ij}\tau^{kl}+\frac{1}{m_{\phi}}\left(K^i\right)^2-\frac{1}{f_{\theta}}\left(S^0\right)^2-\frac{1}{f_u}\left(\tau^{0i}\right)^2-\frac{1}{f_{\phi}}
\left(K^0\right)^2\right.\\
&-2\left(\partial_i\theta\right)S^i-2\left(D_iu_j\right)\tau^{ij}-2\left(D_i\phi\right)K^i   -2\left(\partial_0\theta\right)S^0 - 2\left(\partial_0u_i\right)\tau^{0i} - 2\left(\partial_0\phi\right)K^0\Biggr]\,,
\end{split} 
\label{eq;app:action_HS}
\end{equation}
where
\begin{equation}
\begin{split}
\quad S^i&=m_{\theta}\partial_i\theta\,, \qquad \tau^{ij}=C_{ijkl}(D_ku_l)=C_{ijkl}(\partial_ku_l+\epsilon_{kl}\theta)\,, \qquad K^i=m_{\phi}D_i\phi=m_{\phi}\left(\partial_i\phi+\epsilon_{ij}u_j\right)\,,\\ S^0&= - f_{\theta}\partial_0\theta\,, \hspace{0.55cm} \tau^{0i} = - f_u\partial_0u_i\,, \hspace{4.35cm} K^0 = - f_{\phi}\partial_0\phi\,. 
\end{split}
\end{equation}
A comparison with the definitions of the covariant derivatives in Eq.~(\ref{eq:upper_indices}) allows us to write it more compactly as
\begin{equation}
    S^\mu = \partial^\mu \theta\,,\qquad \tau^{\mu j} = D^\mu u^j\,,\qquad K^\mu = D^\mu \psi\,.
\end{equation}

We include defects in the theory by decomposing $\theta$, $u_i$ and $\phi$ into $\theta=\Bar{\theta}+\theta^{(s)}$, $u_i=\Bar{u}_i+u_i^{(s)}$ and $\phi=\Bar{\phi}+\phi^{(s)}$, respectively, where the bar symbol represents the regular part and the $(s)$ symbol represents the singular part related to the defects. The regular fields appear in the action~(\ref{eq;app:action_HS}) in the following terms:
\begin{equation}
\int d^2xdt\left[-(\partial_\mu\Bar{\theta})S^\mu-(D_\mu\Bar{u}_j)\tau^{\mu j}-(D_\mu\Bar{\phi})K^\mu\right]\,.
\end{equation}
After an integration by parts, the regular fields can be integrated out, producing the 3 equations of motion
\begin{equation}
\partial_\mu S^\mu-\epsilon_{ij}\tau^{ij}=0\,, \hspace{2cm}     
\partial_\mu\tau^{\mu i}+\epsilon_{ij}K^j=0\,, \hspace{2cm} \partial_\mu K^\mu=0\,,
\label{eq:app_3_eoms}
\end{equation}
while the remaining terms constitute the dual action containing the defect contributions
\begin{equation}
S_{\mathrm{dual}}=\frac{1}{2}\int d^2xdt\left[\frac{1}{m_{\theta}}\left(S^i\right)^2+C^{-1}_{ijkl}\tau^{ij}\tau^{kl}+\frac{1}{m_{\phi}}\left(K^i\right)^2-\frac{1}{f_{\theta}}\left(S^0\right)^2-\frac{1}{f_u}\left(\tau^{0i}\right)^2-\frac{1}{f_{\phi}}\left(K^0\right)^2\right]+S_{\mathrm{defects}}.  
\end{equation}

As in the main text, we solve the equations of motion~\eqref{eq:app_3_eoms} by introducing the gauge fields:
\begin{equation}
\begin{split}
S^\mu & = \frac{1}{2}\epsilon^{\mu\nu\rho}F^{(0)}_{\nu\rho}=\epsilon^{\mu\nu\rho}\left(\partial_\nu A^{(0)}_\rho+\delta_{i\nu} A^{(1)}_{i\rho}\right)\,, \\
\tau^{\mu i} &= -\frac{1}{2}\epsilon^{\mu\nu\rho}\epsilon_{ij}F^{(1)}_{j\nu\rho}=-\epsilon^{\mu\nu\rho}\epsilon_{ij}\left(\partial_\nu A^{(1)}_{j\rho}+\delta_{j\nu} A^{(2)}_{\rho}\right)\,, \\
K^\mu &= \frac{1}{2}\epsilon^{\mu\nu\rho}F^{(2)}_{\nu\rho}=\epsilon^{\mu\nu\rho}\partial_\nu A^{(2)}_\rho\,.
\end{split}
\end{equation}
If we split the dual action as $S_{\mathrm{dual}}=S_{\mathrm{EM}}+S_{\mathrm{defects}}$, the electromagnetic counterparts allow us to rewrite $S_{\mathrm{EM}}$ as
\begin{equation}
S_{\mathrm{EM}}=\frac{1}{2}\int d^2xdt\left[\frac{1}{m_{\theta}}\left(F_{i0}^{(0)}\right)^2-\frac{1}{2f_{\theta}}\left(F^{(0)}_{ij}\right)^2+\Tilde{C}_{ijkl}^{-1}F_{ij0}^{(1)}F_{kl0}^{(1)}-\frac{1}{2f_u}\left(F^{(1)}_{ijk}\right)^2+\frac{1}{m_{\phi}}\left(F^{(2)}_{i0}\right)^2-\frac{1}{2f_{\phi}}\left(F_{ij}^{(2)}\right)^2\right]\,.    
\label{app:eq_EM_action}
\end{equation}
In the above we have defined the rotated inverted elasticity tensor $\Tilde{C}_{ijkl}^{-1}=\epsilon_{ia}\epsilon_{jb}\epsilon_{kc}\epsilon_{ld}C_{abcd}^{-1}$; in the case presented in the main text, we have exactly $\Tilde{C}_{ijkl}^{-1}=C_{ijkl}^{-1}$. The electromagnetic action~(\ref{app:eq_EM_action}) represents a quadrupole gauge theory with generic effective coefficients and it becomes equivalent to Eq.~(\ref{eq:effective}) when certain simplifying assumptions are applied to these coefficients. On the other hand, the singular part of the action reads
\begin{equation}
S_{\mathrm{defects}}=\int d^2xdt\left[-\left(\partial_\mu\theta^{(s)}\right)S^\mu-\left(D_\mu u^{(s)}_j\right)\tau^{\mu j}-\left(D_\mu\phi^{(s)}\right)K^\mu\right]\,.
\end{equation}
Performing an integration by parts and grouping the terms corresponding to the different gauge fields results in:
\begin{equation}
S_{\mathrm{defects}}=\int d^2xdt \left(J^{(0)\mu}A^{(0)}_\mu+J^{(1)\mu}_iA^{(1)}_{i\mu}+J^{(2)\mu}A^{(2)}_\mu\right)\,,
\end{equation}
where the defect currents read
\begin{equation}
    J^{(0)\mu}  =-\epsilon^{\mu\nu\lambda}\partial_\nu\partial_\lambda \theta^{(s)}\,, \quad
    J^{(1)\mu}_i = \epsilon^{\mu\nu\lambda}\left(\epsilon_{li}\partial_\nu\partial_\lambda u_l^{(s)} - 2\delta_{i\lambda} \partial_\nu \theta^{(s)}\right)\,, \quad
    J^{(2)\mu} =-\epsilon^{\mu\nu\lambda}\left(\partial_\nu \partial_\lambda\phi^{(s)} + 2\delta_{\lambda k}\epsilon_{kj}\partial_\nu u^{(s)}_j\right)\,.
\end{equation}

\end{document}